\documentclass[pra,aps,twocolumn]{revtex4}

\usepackage{epsfig}

\newcommand{\be}{\begin{equation}}
\newcommand{\ee}{\end{equation}}
\newcommand{\ba}{\begin{eqnarray}}
\newcommand{\ea}{\end{eqnarray}}
\newcommand{\ban}{\begin{eqnarray*}}
\newcommand{\ean}{\end{eqnarray*}}

\newcommand{\braket}[2]{\mbox{$ \langle #1 | #2 \rangle $}}
\newcommand{\moy}[1]{\left\langle #1 \right\rangle}
\newcommand{\sandwich}[3]{\mbox{$ \langle #1 | #2 | #3 \rangle $}}
\newcommand{\ket}[1]{\mbox{$ | #1 \rangle $}}
\newcommand{\bra}[1]{\mbox{$ \langle #1 | $}}
\newcommand{\com}[2]{\left[ #1\,,\,#2 \right]}

\newcommand{\demi}{\frac{1}{2}}

\newcommand{\real}{\begin{picture}(8,8)\put(0,0){R}\put(0,0){\line(0,1){7}}\end{picture}}

\newcommand{\one}{\leavevmode\hbox{\small1\normalsize\kern-.33em1}}

\def\tr{\mbox{tr}}

\begin{document}

\title{Uncertainty Relations for Positive Operator Valued Measures}

\author{Serge Massar}
\affiliation{Laboratoire d'Information Quantique, {C.P.} 225, Universit\'{e} Libre de Bruxelles (U.L.B.), Boulevard du Triomphe, B-1050 Bruxelles, Belgium}

\date{\today}

\begin{abstract}
How much unavoidable randomness is generated by a Positive Operator Valued Measure (POVM)? We address this question using two complementary approaches. First we study the variance of a real variable associated to the POVM outcomes. In this context we introduce an uncertainty operator which measures how much additional noise is introduced by carrying out a POVM rather than a von Neumann measurement. We illustrate this first approach by studying the variances of joint estimates of $\sigma_x$ and $\sigma_z$ for spin $1/2$ particles. We show that for unbiased measurements the sum of these variances is lower bounded by $1$.  In our second approach we study the entropy of the POVM outcomes. In particular we try to establish lower bounds on the entropy of the POVM outcomes. We illustrate this second approach by examples. 
\end{abstract}

\maketitle


\section{Introduction}


Richard Feynman famously said that ``no one understands quantum mechanics". Nevertheless, practitioners of quantum mechanics have a number of mental pictures that  help them reason about the theory. Arguably the most important of them is the uncertainty principle \cite{Heisenberg}, which both gives fundamental insights and provides important quantitative predictions.

The uncertainty principle itself comes in several forms. The best known bounds the variances of two operators $A=\sum_k a_k \ket{a_k}\bra{a_k}$ and $B=\sum_l b_l \ket{b_l}\bra{b_l}$ measured in the quantum state $\ket{\psi}$. 
The resulting inequality \cite{Robertson} is:
 \be
\Delta A \ \Delta B \geq \demi | \sandwich{\psi}{\com{A}{B}}{\psi}|
\label{DADB}
\ee
where $\Delta A^2 = \moy{A^2}-\moy{A}^2$, $\Delta B^2 =  \moy{B^2}-\moy{B}^2$. 
In terms of position and momentum operators it takes the form
\cite{Kennard}:
\be\Delta x \ \Delta p \geq \frac{\hbar}{2}\ .
\label{DxDp}
\ee 

An alternative formulation based on
entropic uncertainty relations \cite{Deutsch}
provides bounds on the entropy of the outcomes of measurements of $A$ and $B$. The version conjectured in \cite{Kraus} and proved in \cite{MU} (for the particular case of position and momentum see \cite{BBM}) takes the form:
 \be
H(A) + H(B) \geq - \log_2 \max_{kl} |\braket{a_k}{b_l}|^2
\label{HAHB}
\ee
where $H(A)=-\sum_k p_A(k) \log_2 p_A(k)$ is the Shannon entropy of the probability distribution $p_A(k)=|\braket{a_k}{\psi}|^2$ and similarly for $H(B)$. 
Entropic uncertainty relations have some conceptual advantages over (\ref{DADB}): they are independent of the values $a_k$, $b_l$ one assigns (often arbitrarily) to the outcomes of the measurements of $A$ and $B$; and their right hand side (rhs) is independent of the quantum state $\ket{\psi}$, 
whereas the rhs of (\ref{DADB}) can vanish even though $\Delta A$ and $\Delta B$ are both positive.
Note that in the above formulations the measurements of $A$ and $B$ are mutually exclusive: they cannot both be carried out. 

A conceptually different form of the uncertainty principle concerns measurements that simultaneously estimate two non commuting operators $A$ and $B$. The precision $\Delta A$ and $\Delta B$ with which $A$ and $B$ are jointly estimated should obey a constraint similar to eq. (\ref{DADB}).
But proving such constraints is difficult, see e.g. \cite{WZB,ETh,OKiukas,Werner,BHL,Breview,Englert,Durr,Hall07,Hall04}.

Simultaneously estimating two observables cannot be carried out within the usual framework of Projector Valued Measures (PVM). Rather it must be formulated within the more general context of 
 Positive Operator Valued Measures (POVM). Formally a POVM  ${\cal M}$ is described by a set of positive operators that sum to the identity: ${\cal M} = \{ m_k \}$, $m_k\geq 0$, $\sum_k m_k = \one$.

POVM's play an essential role in measurement theory as they describe measurements affected by noise, fuzzy measurements, measurements that simultaneously estimate two observables, etc...  
POVM's also play an essential role in quantum information: they are often the measurements which allow the most information to be extracted from a quantum system\cite{PeresWootters,MassarPopescu}, they are widely used for quantum communication tasks, etc... However POVM's suffer from the fact that (except when some of the POVM elements are projectors) there is no state $\ket{\psi}$ for which the outcome of the POVM is fixed. That is, the outcomes of POVM's are affected by unavoidable noise. This raises a fundamental question: {\em why is it that, given their unavoidable noise, POVM's are nevertheless sometimes better than PVM's for information processing tasks?} 
A partial explanation is that the noise of a POVM is more uniformly distributed over Hilbert space: there are many states on which a PVM will give completely random results, whereas many POVM's never give completely random outcomes. But it is unclear whether this is the whole answer.
Indeed there at least one very specific context in which the randomness produced by a POVM can be removed: if the same POVM must be carried out on many independent states, then one can devise a collective POVM acting on all the states which is almost a PVM, and which when restricted to a single system, acts as the original POVM\cite{foot}. 

The inherent randomness of POVM's may also explain some of their limitations. For instance there is to our knowledge no non locality experiment for which POVM's are better than PVM's, and this is probably due to the added noise coming from POVM's. Better understanding the potentialities and limitations of POVM's thus hinges on better understanding the unavoidable noise they add to experiments. In the present work we address this task, and carry out {\em a systematic study of the noise produced by POVM's.}
There are already many works that address aspects of this question \cite{WZB,ETh,OKiukas,Werner,BHL,Breview,Englert,Durr}, often in the context of joint measurements of position and momentum. 
 But it is clear that a unified approach is necessary that does not focus on the technical difficulties of infinite dimensional spaces, but rather on the conceptual issues involved. The present work aims to fill this gap by following two complementary approaches,  similar in spirit to the complementary approaches  provided by  the Robertson and entropic uncertainty relations mentioned above. 

\section{Uncertainty Operator}\label{uncertaintyse}


In our first approach (see also \cite{ETh,BHL,Hall07,Hall04})  we suppose that to each POVM element $m_k$ one associates a real value $\mu_k \in \real$. This association of course suffers from the same limitations as the Robertson inequality: the choice of  the $\mu_k$ is to some extent arbitrary. Different choices will yield different estimates of the uncertainty. Furthermore, it is often natural to associate several values to the same POVM element $m_k$. We will then have several uncertainties associated with the same POVM.

The expectation value of $\mu_k$ is
\be
\bar \mu = \sum_k \mu_k \sandwich \psi { m_k} \psi = \sandwich \psi { M} \psi
\ee
where we introduce the operator
\be
M=\sum_k \mu_k m_k \ .
\ee
The variance of $\mu_k$ is
\ba
Var ( \mu ) &=&   \sum_k \mu_k^2 \sandwich \psi { m_k} \psi - {\bar \mu}^2 \label{V1}\\
&=& \sandwich \psi { M^2 - {\bar \mu}^2 } \psi + \sandwich \psi { \Delta M^2 } \psi 
\label{V2}
\ea
where we introduce the uncertainty operator
\be
\Delta M^2 = \sum_k \mu_k^2 m_k - M^2
\ee
The first term in eq. (\ref{V2}) is the variance of $\mu_k$ which would arise if one was to ``measure the observable $M$" in the usual sense. Note that there are always some states $\ket{\psi}$ for which this term vanishes. The second term is the additional uncertainty which arises because one is measuring a POVM and not a PVM. It does not depend on the average value $\bar \mu$. The uncertainty operator $ \Delta M^2 $ thus characterizes the extra noise coming from the POVM. It has the following important properties:\\
\begin{enumerate}
\item {\em Positivity:} ${\Delta M^2}  \geq 0$ is a positive operator, which follows from rewriting it in the manifestly positive 
form $\Delta M^2 = \sum_k (M- \mu_k) m_k (M-\mu_k)$ \cite{Hall07};\\
\item {\em Vanishing on PVM's:} ${\Delta M^2} =0$ if ${\cal M}$ is a PVM;\\
\item {\em Strict positivity on POVM's:} if ${\cal M}$ is not a PVM, then there exists a choice of $\mu_k$ such that $ {\Delta M^2} >0 $ is strictly positive (simply take $\mu_k = \delta_{k k_0}$ to be zero except for one value $k_0$, with $k_0$ such that $m_{k_0}$ is not a projector);\\
\item {\em Reduction to classical random variables:} if the POVM elements are all proportional to the identity 
$m_k =  |m_k| \one$, then the probabilities of the outcomes are independent of the quantum state, and $\mu_k$ is a classical random variable, with associated probabilities $p_k=|m_k|/d$ where $d$ is the dimension of the Hilbert space. In this case the first term in eq. (\ref{V2}) vanishes and ${\Delta M^2} = \one \sum_k (\mu_k - \bar \mu)^2 p_k$ is the variance of the classical random variable $\mu_k$.\\
\item {\em Additivity under tensor product:} Consider two POVM's and their associated values $\{m_i,\mu_i\}$ and $\{n_j , \nu_j\}$ acting on different systems. We construct the tensor product POVM as $\{m_i\otimes n_j , \mu_i+\nu_j\}$, where we associate to each outcome $(i,j)$ the sum of the values $\mu_i + \nu_j$ for each outcome. Then the uncertainty operator for the tensor product POVM is
$\Delta (M\otimes N)^2 = \Delta M^2\otimes \one + \one \otimes \Delta N^2$.\\
\item {\em Increase under convex combination:} Consider two POVM's and their associated values $\{m_i,\mu_i\}$ and $\{n_j , \nu_j\}$ acting on the same system. We construct the convex combination of these two POVM's by realizing the first POVM with probability $p$, and the second POVM with probability $q$ ($p+q=1$) to obtain a POVM with elements $p m_i$ and $q n_j$ to which are associated the values $\mu_i$ and $\nu_j$ respectively. The uncertainty operator for the convex combination POVM is
$\Delta (pM+ qN)^2 =p \Delta M^2 + q \Delta N^2   + pq (M-N)^2$.
\end{enumerate}

Properties 5 and 6 are in direct analogy with the way the variances of  independent classical random variables behave under addition and convex combination.

Following the ideas developped in \cite{Hall07}, see also \cite{Hall04}, some further properties of the uncertainty operator can be obtained by
introducing Naimark's extension of the POVM ${\cal M}$.
Naimark's theorem states that there exists an extended Hilbert space $\tilde H$ which is the direct sum of the system Hilbert space $H$ (on which ${\cal M}$ acts) and an ancillary Hilbert space $H'$ ($\tilde H = H \oplus H'$); and a basis of the extended Hilbert space $\ket{\tilde m_k}$ ($ \braket{\tilde m_{l}}{\tilde m_k} =  \delta_{k l}$) which restricted to the system space gives the POVM ${\cal M}$: 
$\ket{\tilde m_k}=\ket{m_k} + \ket{m'_k}$,
 where
$\ket{m_k} \in H$ and $\ket{m'_k} \in H'$. (For simplicity we suppose that the POVM elements are rank 1: $m_k = \ket{m_k}\bra{m_k}$, the general case follows by taking several of the $\mu_k$ to have the same value).

We can now then define a Hermitian operator on the extended space as $M_E=\sum_k \mu_k  \ket {\tilde m_k}\bra{\tilde m_k}$.
Then, if we denote by $P$ the projector onto the system space $H$, we can rewrite the uncertainty operator as
$\Delta M^2=P (M_E - M)^2 P$ which shows that the uncertainty operator measures the difference between $M$ and any Naimark extension of the POVM ${\cal M}$.
More generally, it is possible to introduce a notion of distance between the POVM ${\cal M}$ and Hermitian operators $A$ acting on $H$ by  $d(A, {\cal M})^2= \tr [ P(M_E- A)^2 P]$. The Hermitian operator which minimises this distance is $M$, and the minimum distance is 
$d(M, {\cal M})^2 = \tr \Delta M^2$. We can also consider the statistical distance between $M_E$ and Hermitian operators $A$ acting on $H$:
$\langle \psi | (M_E- A)^2 | \psi \rangle$ where $|\psi\rangle \in H$ is a state in the system Hilbert space. Once more the statistical distance is minimised if we take $A=M$, and the minimal statistical distance is $\langle \psi | \Delta M^2 |\psi \rangle$.

Finally it is interesting to note that the uncertainty relation eq. (\ref{V2})
can also be derived from the Robertson inequality. To this end recall that the Naimark extension is not unique:
thus for instance $\ket{\tilde m_k}=\ket{m_k} + \ket{m'_k}$ and $\ket{\bar m_k}=\ket{m_k} + i \ket{m'_k}$ are two valid Naimark extensions of the same POVM ${\cal M}$. Hence we can define two operators $\tilde M=\sum_k \mu_k \ket{\tilde m_k}\bra{\tilde m_k}$ and
$\bar M=\sum_k \mu_k \ket{\bar m_k}\bra{\bar m_k}$. Applying the Robertson inequality eq. (\ref{DADB}) we have
$\Delta \tilde M \Delta \bar M \geq \frac{1}{2} |\langle [\tilde M,\bar M]\rangle|$. But if the quantum state $\ket \psi$ has support only in the system space $H$, 
then $\Delta \tilde M =\Delta \bar M=Var(\mu)^{1/2}$ and $\frac{1}{2}[\tilde M,\bar M] = i\Delta M^2$, yielding the second term in eq. (\ref{V2}). (If we add on the right hand side of the Robertson inequality a term containing the anti commutator of the two operators, we also recover the first term in eq. (\ref{V2})).

\section{Applications}

\subsection{Joint Measurements of Position and Momentum}


A particularly 
well studied example (for more details see \cite{BHL} and references therein) is the  covariant joint measurement of 
$p$ and $x$ of the form
$m_{px}=D_{px} \frac{ m_{0} }{2 \pi} D_{px}^\dagger$ where $D_{px}$ is the displacement operator in phase space  and $m_0$ is a normalized state. 

If one uses $m_{pq}$ to estimate $q$, then one will associate with $m_{pq}$ the value $q$. In this case one finds that the associated operator 
$Q = \int dp dq \ q \hat m_{pq} = \hat q -  \one \left ( \int dq q \ \sandwich{q}{m_{0}}{q} \right)$ is just the operator $q$ displaced to coincide with the center of $m_0$, and the uncertainty operator is just the identity times the variance of $q$ in the state $m_0$:
$\hat {\Delta Q^2} = \one \Delta q^2_{m_0}$
with $\Delta q^2_{m_0}
=\left ( 
\int dq q^2 \ \sandwich{q}{m_{0}}{q} -
\left(\int dq q \ \sandwich{q}{m_{0}}{q} \right)^2 \right)$.

From eq. (\ref{V2}) it then follows that the variance of this estimator of $q$ in state $\ket{\psi}$ is the sum of the variances of $\hat q$ in states $\ket{\psi}$ and $m_0$:
$\Delta q^2 = \Delta q^2_{\psi} + \Delta q^2_{m_0}$.
Similarly, if one associates with $m_{pq}$ the value of $p$, one finds that the variance of this estimator of $p$ in state $\ket{\psi}$ is
$\Delta p^2 = \Delta p^2_{\psi} + \Delta p^2_{m_0}$. Putting these two results together and using eq. (\ref{DxDp}), we obtain a bound on the variances of the joint estimates $p$ and $q$:
$\Delta q^2 \Delta p^2 \geq \Delta q^2_\psi \Delta p^2_\psi +\Delta q^2_{m_0} \Delta p^2_{m_0}
\geq
\hbar$
which is twice the variance of $q$ and $p$ separately.

\subsection{Joint Estimates of $\sigma_x$ and $\sigma_z$}

Another illustration of the power of our approach is provided by the problem of jointly estimating both the $\sigma_z$  and $\sigma_x$ observables of a spin $1/2$ particle.
Such a measurement will be described by a POVM with 4 outcomes: $m_{zx}$,  where the label $z=\pm 1$ ($x=\pm 1$) corresponds to inferring that a measurement of $\sigma_z$ ($\sigma_x$) would preferentially have given the $z=+1$ or $z=-1$ ($x=+1$ or $x=-1$) outcome.

Let us first illustrate such a POVM by the following example: 
\be
m_{zx}=\frac{\one}{4} + z \frac{\cos \theta}{4}\sigma_z + x \frac{\sin \theta}{4}\sigma_x\ .
\label{mzx}\ee
 If we use this POVM to estimate the $z$ ($x$) component of the spin, then the associated operators are 
$Z=\sum_{z,x=\pm 1} z m_{zx} = \cos \theta \sigma_z$ and $X=\sum_{z,x=\pm 1} x m_{zx} = \sin \theta \sigma_x$. Thus one is indeed simultaneously estimating the $z$ and $x$ components of the spin, but with reduced sensitivity with respect to measuring the observables $\sigma_z$ and $\sigma_x$ separately. The associated uncertainty operators are
\ba
\Delta Z^2 &=& \sum_{zx} z^2 m_{zx} - (\sum_{zx} z m_{zx})^2 = (1 -\cos^2\theta) \one \quad \nonumber\\
 \Delta X^2 &=& \sum_{zx} x^2 m_{zx} - (\sum_{zx} x m_{zx})^2 =(1 -\sin^2\theta) \one \quad 
 \label{exampleZX} \ea
which we can group in the relation $\Delta X^2  + \Delta Z^2 =\one$
which implies  the constraint on the sum of the variances 
\be
Var (Z) + Var (X) \geq 1\ .
\label{VVVV}\ee 
This constraint has been derived previously for the particular POVM's eq. (\ref{mzx}), see \cite{Breview} and references therein.

Let us now generalise eq. (\ref{VVVV})  to arbitrary POVM's with 4 outcomes $z,x=\pm 1$, where outcome $z=+1,x=+1$ corresponds to guessing that measurements of $\sigma_z$ and of $\sigma_x$ would both have given outcome $+1$; and similarly for the other values of $z,x$. We require that the POVM be {\em unbiased} in the following weak sense: we require that for the completely mixed state $\rho=\one/2$, the probabilities of the different outcomes $z,x$ are all equally probable. This implies that the POVM elements can be written as
$m_{zx}=\one/4 + \vec v_{zx} \cdot \vec\sigma$ with $|\vec v_{zx}|\leq 1/4$ and $\sum_{zx} \vec v_{zx} = 0$.
In addition for the measurement to be {\em faithfull} we require that $\vec v_{++}$ has positive $z$ and $x$ components. This corresponds to requiring that the state $\psi$ that maximizes the probability of getting outcome $++$ would give with high probability $\sigma_z=+1$ and $\sigma_x =+1$ if one were to measure these operators. Similar conditions hold for the other values of $z$ and $x$.
As before we associate with outcome $z,x$ the values $Z=z, X=x$. It then follows (in fact from the unbiasedness condition alone) that 
\be 
\Delta Z^2 + \Delta X^2 = \one (2 - 2|\vec v_{++} - \vec v_{--}|^2 - 2|\vec v_{+-} - \vec v_{-+}|^2) \geq \one \label{DDD}\ee
which implies eq. (\ref{VVVV}).
Note also that equality in eq. (\ref{VVVV}) is obtained if and only if
$v_{++} = - \vec v_{--}$, $\vec v_{+-} =- \vec v_{-+}$ and $|\vec v_{zx}|=1/4$. Thus the class of unbiased POVM which jointly estimates $\sigma_z$ and $\sigma_x$ with the least added noise contains, but is larger, than the example eq. (\ref{exampleZX}). (Note that all these optimal POVM's can be realized by measuring with probability $1/2$ the operator $\vec v_{++} \cdot \vec \sigma$ and with probability $1/2$ the operator $\vec v_{+-} \cdot \vec \sigma$).

The above problem of jointly measuring $\sigma_x$ and $\sigma_z$ for a spin $1/2$ particle is related to, and motivated by, the 
problem of better understanding uncertainty relations for Mach-Zehnder interferometers.
Indeed in the case of a Mach-Zehnder interferometer there are two natural questions to ask: through which arm of the interferometer does the particle pass, and by which output port does it exit. Because the Mach-Zehnder interferometer is a two dimensional system, there is a formal mapping to the spin $1/2$ case, with the path observable associated to $\sigma_z$ and the output port observable associated with $\sigma_x$.

Two uncertainty relations have been derived for the Mach Zehnder problem. The first is similar in spirit to the Robertson inequality, as it relates the variances of two incompatible observables measured on the same state $\psi$. It relates the Predictability ${\cal P}$ of the path taken by the particle to the Visibility ${\cal V}$ of the interference fringes by ${\cal P} + {\cal V} \leq 1$\cite{GY,V}. 
We have  the following correspondence with the spin $1/2$ case: ${\cal P}=1-Var(\sigma_z)_\psi$ and ${\cal V}\geq 1-Var(\sigma_x)_\psi$. (The inequality in the last relation comes from the fact that the visibility is defined by the maximum contrast when the phase in the interferometer is varied. The maximum contrast may not occur for a measurement of $\sigma_x$, but of $\cos \theta \sigma_x + \sin \theta \sigma_y$ for some $\theta$).

The second uncertainty relation concerns {\em joint estimates} of the path a particle takes in a Mach-Zehnder interferometer and the output port by which it exits\cite{Englert} (see also \cite{ESW,Breview}). The approach  adopted in these works is to attach to the particle a which way marker which registers information about which path the particle took in the interferometer. One then measures both the marker and the output port, thereby implementing a POVM with four outputs (see \cite{DNR,SKE,Obrien} for experimental realisations of this idea). Englert \cite{Englert} obtained in this context an uncertainty relation which constrains the liklehood of correctly guessing the path 
taken by the particle (the distinguishability ${\cal D}$) and the visibility ${\cal V}$ of the interference fringes) in a joint measurement using a which way marker: ${\cal D} + {\cal V} \leq 1$.

Our result eq. (\ref{VVVV}) is related to the work of Englert, since it constrains the precision with which one can {\em jointly estimate} the path and output port taken by the particle. But here we have adopted a more abstract approach, and consider arbitrary unbiased POVM's with 4 outcomes, rather than specific implementations using which way markers.

\section{Entropic uncertainty relation for POVM's}\label{entropic}

An alternative approach toward understanding the amount of randomness generated by a POVM is to lower bound  entropy of the outcomes $H({\cal M})=-\sum_k p(k) \log_2 p(k)$ with $p(k) = \sandwich{\psi}{m_k}{\psi}$. A very simple bound of this type is provided by the largest eigenvalue of the POVM elements $m_k$:
\be
H({\cal M}) \geq -\log_2 \left(\max_{k \psi} \sandwich{\psi}{m_k}{\psi}\right)\ .
\label{HM1}
\ee
Thus, except in the trivial case when one of the POVM elements is a projector, the entropy of the POVM outcomes is always positive. In some cases we have been able to improve this bound.

When the POVM can be realized by carrying out with probability $1/2$ one of two non degenerate PVM's
$\{ \ket{a_k}\}$ and $\{ \ket{b_l} \}$ (with $\braket{a_k}{a_{k'}}=\delta_{kk'}$ and $\braket{b_l}{b_{l'}}=\delta_{ll'}$) 
then eq. (\ref{HAHB}) implies
$H({\cal M}) \geq 1 - \frac{1}{2} \log_2 \max_{k,l} |\braket{a_k}{b_l}|^2 $.
Applied to the POVM eq. (\ref{mzx}) this yields the bound
\be H(\{m_{zx}\})\geq 1 -\log_2 \cos \theta \quad , \quad 0 \leq \theta \leq \pi/4 \label{Mmzx}\ee
(which is tight when $\theta = 0$ or $\theta = \pi/4$ but is suboptimal in between). 
From eq. (\ref{Mmzx}) it follows that as the POVM $\{m_{zx}\}$ goes from estimating only $\sigma_z$ ($\theta = 0$) to estimating $\sigma_z$ and $\sigma_x$ with equal sensitivity ($\theta = \pi/4$), the minimum entropy generated by the POVM increases from $1$ to $3/2$ bits. This is a particular example of what we expect is a general trade off: POVM's that probe more uniformly the Hilbert space (which can be a useful property for information processing) generate more randomness (a deleterious property).

We have also improved on eq. (\ref{HM1}) using a different method. Let us consider bounds on the entropies of two POVM's, $ {\cal M } = \{ \ket{m_k}\bra{m_k} \}$ and 
$ {\cal N } = \{ \ket{n_l}\bra{n_l} \}$, whose elements are all rank 1, acting on the same state. As was noted in \cite{H97}, the proof of the entropic uncertainty relation given in \cite{MU} immediately generalizes to:
\be
H({\cal M})+H({\cal N}) \geq -\log_2 \max_{k l} | \braket{m_k}{n_l}| \ .
\label{HHH}
\ee
This can be strengthened by noting that the Naimark extension is not unique: any PVM of the form
$U'\ket{\tilde m_k}=\ket{m_k} + U'\ket{m'_k}$ with $U'$ acting only on the ancillary Hilbert space $H'$ is a possible extension of the POVM ${\cal M}$. Applying eq. (\ref{HAHB}) to the Naimark extension of the POVM's ${\cal M}$ and ${\cal N}$ and taking the best such bound yields:
\be
H(M)+H(N) \geq \max_{ U'} -\log_2  \max_{k l} | \sandwich{\tilde m_k}{U'}{\tilde n_l}| \ .
\label{HMHN}
\ee

We can now go back to the case of a single POVM. Let us take in eq. (\ref{HMHN})  the two POVM's to be identical but with different Naimark extensions. This yields the entropic uncertainty relation for a single POVM:
\be
H({\cal M}) \geq \max_{U'} -\frac{1}{2}\log_2 \max_{k l } | \sandwich{\tilde m_k}{U'}{\tilde m_l}| \ .
\label{HM}
\ee

As an illustration we consider the POVM  described in \cite{MassarPopescu} which  acts on the symmetric space of two spin 1/2 particles and is composed of 4 elements, each proportional to the projector onto two parallel spins oriented along the 4 corners of a tetrahedron. The Naimark extension of this POVM can be constructed explicitly as $\ket{\tilde m_0}=\frac{\sqrt{3}}{2}\ket{\uparrow_z}\ket{\uparrow_z}+\frac{1}{2}\ket{a}$,
$\ket{\tilde m_j}=\frac{\sqrt{3}}{2}\ket{\uparrow_j}\ket{\uparrow_j}-\frac{1}{2}\ket{a}$ where
$\ket{\uparrow_j}=(\ket{\uparrow_z} + \sqrt{2}e^{i 2 \pi j /3} \ket{\downarrow_z})/\sqrt{3}$, $j=1,2,3$.
If we take $U'\ket{a}=-\ket{a}$ (with $U'$ acting as the identity on the space of the two spins), then 
eq. (\ref{HM}) implies that the  entropy of this POVM is bounded by $H({\cal M})\geq 1$ bit, which is significantly better than the bound $H({\cal M})\geq -\log_2 (3/4)$ which follows from eq. (\ref{HM1}).

\section{Conclusion}

In summary, in this paper we have studied how much unavoidable randomness is generated by a Positive Operator Valued Measure (POVM). We address this question using two complementary approaches. First we study the variance of a variable associated to the POVM outcomes. We illustrate this method by establishing an inequality on the variances of joint estimates of $\sigma_x$ and $\sigma_z$ for spin $1/2$ particles. Second we study lower bounds on the entropy of the POVM outcomes. We also illustrate this second approach by examples.


{\bf Acknowledgements}
I would like to thank Jeremy O'Brien, Paul Bush, and specially Michael Hall, for comments on the first version of this manuscript. 
I acknowledge support by EU project QAP contract 015848, and by the IAP project -Belgium Science Policy- P6/10 Photonics@be.


\end{document}